\documentclass[doublecol]{epl2}
\usepackage{amsmath,amssymb,graphicx}

\title{ Transport of interacting electrons through a potential
barrier:\\
nonperturbative RG approach}
\shorttitle{Transport of interacting electrons through a potential
barrier}

\author{D.N. Aristov \inst{1,2}\thanks{%
On leave from Petersburg Nuclear Physics Institute, Gatchina 188300,
Russia}
\and P. W\"{o}lfle \inst{1,2,3} }
\shortauthor{D.N. Aristov \etal}

\institute{
  \inst{1} Institut f\"ur Theorie der Kondensierten Materie,
Universit\"at Karlsruhe, 76128 Karlsruhe, Germany \\
  \inst{2} Center for Functional Nanostructures, Universit\"at
Karlsruhe, 76128 Karlsruhe, Germany\\
\inst{3} Institut f\"ur Nanotechnologie, Forschungszentrum
Karlsruhe, 76021 Karlsruhe, Germany
}
\pacs{71.10.Pm}{Fermions in reduced dimensions (anyons, composite
           fermions, Luttinger liquid, etc.)}
\pacs{73.63.Nm}{Quantum wires (Electronic transport in nanoscale
           materials and structures)}
\pacs{71.10.-w}{Theories and models of many-electron systems}
\abstract{
We calculate the linear response conductance of electrons in a Luttinger
liquid with arbitrary interaction $g_{2}$, and subject to a potential
barrier of arbitrary strength, as a function of temperature. We first map the Hamiltonian in the
basis of scattering states into an effective low energy Hamiltonian in
current algebra form. Analyzing the perturbation theory in the fermionic
representation the diagrams contributing to the renormalization group
(RG) $\beta$-function are identified. A universal part of the
$\beta$-function is given by a ladder series and summed
to all orders in $g_{2}$. First non-universal corrections beyond the
ladder series are discussed. The RG-equation for the temperature
dependent conductance is solved analytically. Our result
agrees with known limiting cases.}

\begin{document}

\maketitle

Electron transport in nanowires has been studied theoretically for more than
20 years. Initially it was found that electron-electron interaction affects
even the conductance of a clean wire \cite{Apel, Kane} . In the case of
realistic boundary conditions, namely attaching ideal leads to the
interacting quantum wire, the two-point conductance of a clean wire is that
of the leads, equal to one conductance quantum per channel, irrespective of
the (forward scattering) interaction \cite{Maslov,Safi}. Alternatively, it
has been argued that the screening of the external field by the interacting
electron system leads to a renormalization of the conductance to its ideal
value of unity \cite{Oreg}. The work of Kane and Fisher \cite{Kane} and
Furusaki and Nagaosa \cite{Furusaki} showed, that interaction has a dramatic
effect on the conductance in the presence of a potential barrier. For
repulsive interaction these authors found that the conductance tends to zero
as the temperature, $T$, or more generally, the excitation energy of the
electrons approaches zero. This was shown at low temperature in the limits
of weak potential barrier and strong potential barrier (tunneling limit),
and
 {at special values of the interaction parameter,
$K=\frac{1}{2}$
and $K=\frac{1}{3}$, for all temperatures \cite{Kane,Weiss,Fendley95}. }
It has
been argued, that these results apply to a contact free four-point
measurement, which is best realized by measuring the absorption losses
of an a.c. field in the limit $\omega \rightarrow 0$ \cite{Maslov,Safi}
(for a different view, see \cite{Oreg}).

We recall that the reason for the strong suppression of the conductance by a
repulsive interaction is that the Friedel oscillations of the charge density
around the potential barrier act as a spatially increasingly extended
effective potential as the temperature is lowered. A proper treatment of the
two-point conductance in the limit of weak interaction, taking into account
the gradual build-up of the Friedel oscillations as the infrared cutoff is
lowered has been given by Yue, Matveev and Glazman \cite{Yue}. These authors
used the perturbative RG for fermions to derive the conductance for an
arbitrary (but short) potential barrier. A generalization of that approach
to the case of two barriers has been given in \cite{Gornyi}.

In this letter we propose to extend the approach of Yue et al. to arbitrary
strength of interaction. We argue that the $\beta $-function of the
RG-equation for the conductance can be obtained in very good approximation
by summing a class of contributions in all orders of the interaction. As we
show below, this is possible in this case, at least at low temperature,
since the class of diagrams with the maximum number of loops in any order,
which are the ones contributing to the $\beta $-function, form a ladder
series. At intermediate temperatures additional diagrams contribute small
corrections to the $\beta$-function. The result of the solution of the
RG-equation for the conductance, using the approximate $\beta$-function thus
obtained, is found to agree with all known results on the scaling behavior
of the conductance, including the case $K=\frac{1}{2}$, but goes far beyond:
it is valid for any interaction strength $K$ and any potential scattering
strength. A careful analysis of the dependence of the result on the cutoff
procedure chosen shows that certain terms in the $\beta $-function are not
universal and depend on the cutoff scheme.

\textit{The Model}. We consider a one-dimensional system (coordinate $x$) of
spinless electrons subject to a potential barrier at $x=0$. The barrier is
characterized by transmission and reflection amplitudes $t=\tilde{t}=\cos
\theta $, $r=-\tilde{r}^{\ast }=i\sin \theta e^{i\phi}$ with negligible
energy dependence in the energy range of interest (width of order of
temperature, $T$, around the Fermi energy, $\epsilon _{F}$). We assume the
extension of the barrier, $a$, to be narrow, $ak_{F}\ll 1$, and neglect all
interaction processes both close to the barrier $|x|<a$ (here $k_{F}$ is the
Fermi wave number) and in the leads, $|x|>L$. Beyond the large scale $L$ the
system is assumed non-interacting, which allows for an asymptotic
single-particle scattering states representation.

If $c_{1k}^{+}$ and $c_{2k}^{+}$ are operators creating electrons in
right-moving and left-moving single particle scattering states of the
barrier ($k > 0$), we may define the electron creation operator $\psi^{+}(x)$
as
\begin{eqnarray}
\psi ^{+}(x) &=&\left\{ \Theta (-x)\left[ \psi _{1}^{+}(x)+r\psi
_{1}^{+}(-x)+\tilde{t}\psi _{2}^{+}(x)\right] \right.  \notag \\
&&+\left. \Theta (x)\left[ t\psi _{1}^{+}(x)+\tilde{r}\psi
_{2}^{+}(-x)+\psi
_{2}^{+}(x)\right] \right\} \,,  \label{1}
\end{eqnarray}
where $\psi _{1,2}^{+}(x)=\int_{0}^{\infty }\frac{\upd k}{2\pi }e^{\pm
ikx}c_{1,2k}^{+}$.

It is convenient to define current operators $J_{\mu }(x)$, $\mu
=0,1,2,3$
by $J_{\mu }(x)=\frac{1}{2}\Sigma _{\alpha ,\beta =1,2}\psi _{\alpha
}^{+}(\alpha x)\tau _{\alpha \beta }^{\mu }\psi _{\beta }(\beta x)$,
where $%
\tau ^{\mu }$ are the Pauli matrices plus the unit matrix, $\psi
_{\alpha
}^{+}(\alpha x)=\psi _{1}^{+}(x)$ or $\psi _{2}^{+}(-x)$ for $\alpha
=1,2$.
We call $J_{0}$ the isocharge current and the vector $\vec{J}%
=(J_{1},J_{2},J_{3})$ the isospin current; these operators obey $U(1)$
and $%
SU(2)$ Kac-Moody algebras, respectively \cite{GoNeTs,AffLud91}. In this
paper we will not make use of these relations, but instead will work in
the
fermion representation. Nonetheless, the representation allows one to
work
in the chiral fermion representation.

The particle density operators for incoming $(i)$ and outgoing $(o)$
particles in terms of the $J_{\mu }$'s are given by (here $x>0$ )
\begin{eqnarray}
\rho _{iR,L}(\mp x) &=&\psi _{1,2}^{+}(\mp x)\psi _{1,2}(\mp
x)=J_{0}(-x)\pm
J_{3}(-x)\,,  \notag \\
\rho _{oR}(x) &=&\left[ t\psi _{1}^{+}(x)+\tilde{r}\psi
_{2}^{+}(-x)\right] %
\left[ t^{\ast }\psi _{1}(x)+\tilde{r}^{\ast }\psi _{2}(-x)\right]
\notag \\
&=&J_{0}(x)+\tilde{J}_{3}(x)  \notag \\
\rho _{oL}(-x) &=&\left[ r\psi _{1}^{+}(x)+\tilde{t}\psi _{2}(-x)\right]
\left[ r^{\ast }\psi _{1}(x)+\tilde{t}^{\ast }\psi _{2}(-x)\right]
\notag \\
&=&J_{0}(x)-\tilde{J}_{3}(x) \,.  \label{2}
\end{eqnarray}
Here $\tilde{J}_{3}=(R\vec{J})_{3}$ is the third component of the
isospin
current vector $\vec{J}$ rotated by the orthogonal matrix $R_{\mu \nu }$
with $R_{33}=|t|^{2}-|r|^{2}=\cos 2\theta $, $R_{32}=\mbox{Im}
\{t\widetilde{%
r}^\ast+ \widetilde{t}r^\ast \} =-\sin 2\theta \cos \phi $,
$R_{31}=\mbox{Re}%
\{t\widetilde{r}^\ast-\widetilde{t}r^\ast \} =\sin 2\theta \sin \phi $.

\begin{figure}[tb]
(a)\includegraphics[width=8cm]{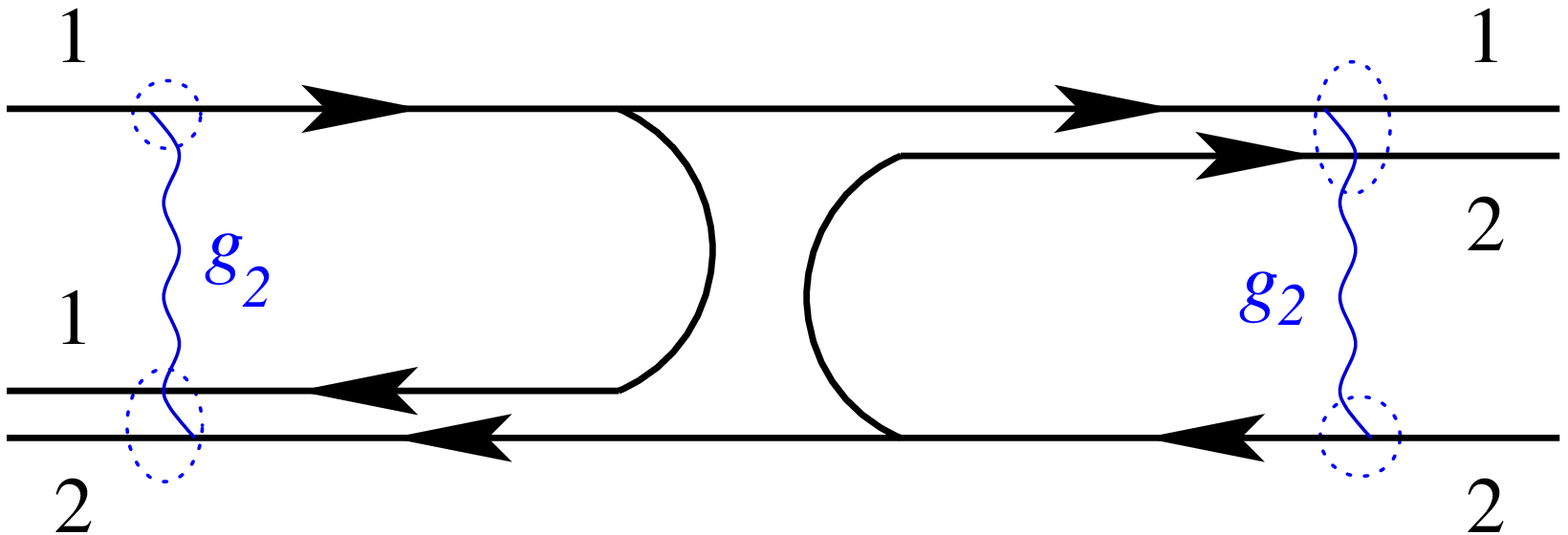} (b)%
\includegraphics[width=8cm]{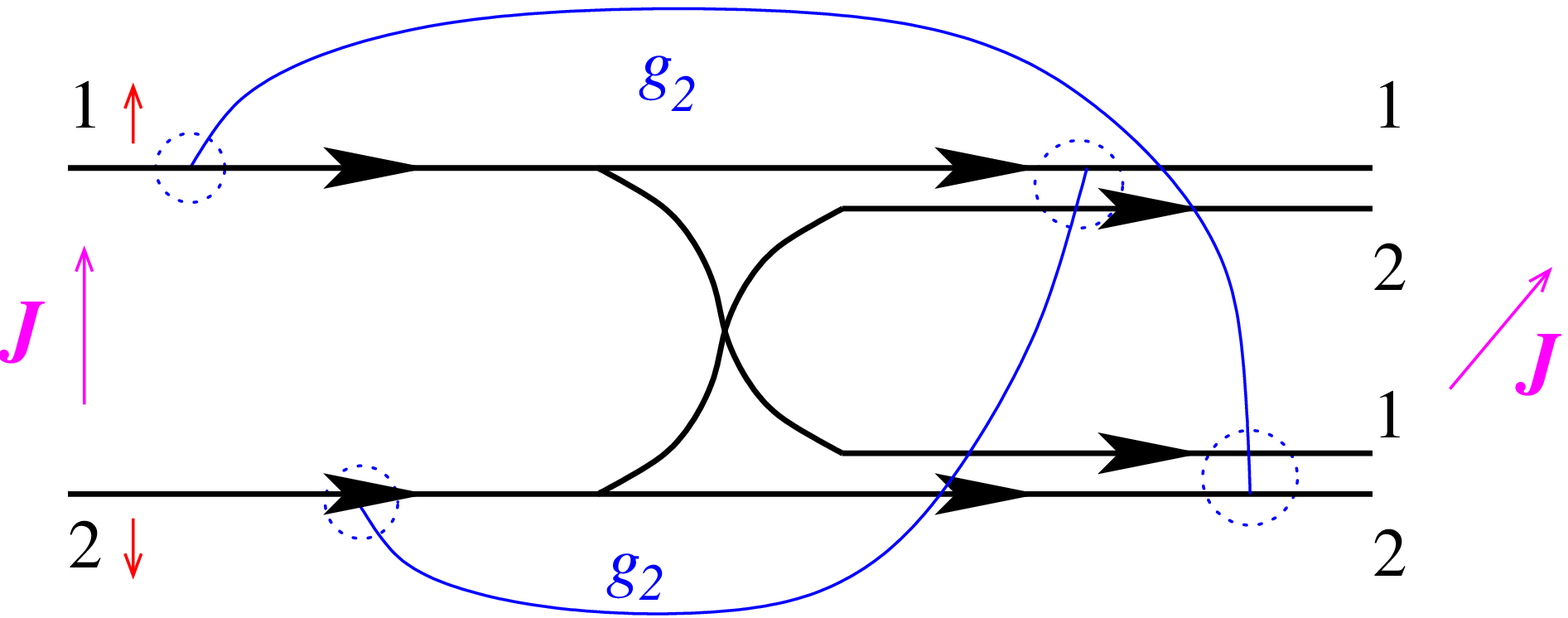}
\caption{ (a)The interaction of the left- and right-going densities in
the
basis of scattered waves. (b) The interaction of the left- and
right-going
densities in the chiral basis, corresponding to non-local interaction,
Eq.\ (\protect\ref{3}). }
\label{fig:scattering}
\end{figure}

We consider a model with interaction constant $g_{2}$ (no
backscattering, no
Umklapp processes). The Hamiltonian is given by $H=H_{0}+H_{1}$, with
\begin{eqnarray}
H_{0} &=&2\pi v_{F}\int_{0}^{\infty }\upd x\left[
J_{0}^{2}(-x)+J_{0}^{2}(x)+J_{3}^{2}(-x)+\tilde{J}_{3}^{2}(x)\right] \,,
\notag \\
H_{1} &=&2g_{2}\int_{a}^{L}\upd x\left[
J_{0}(-x)J_{0}(x)-J_{3}(-x)\tilde{J}%
_{3}(x)\right] \,,  \label{3}
\end{eqnarray}%
where $v_{F}$ is the Fermi velocity. The forward scattering interaction
$g_{4}$ of like-movers may be absorbed into redefinition of $v_{F}$ in
the
usual way \cite{GoNeTs}. In Fig.\ \ref{fig:scattering} we show in a
pictorial way our parametrization of the fermionic densities and the
interaction, $H_{1}$. In Fig.\ \ref{fig:scattering}a the
$g_{2}$-interaction
processes are shown in the usual scattering configuration. The
representation in terms of the currents $J_{\mu }$ is in the chiral
basis
(all particles moving to the right, see Fig.\ \ref{fig:scattering}b),
which
leads to a seemingly nonlocal interaction. It should be clear that
electrons
in the left half space ($x<0$) are not affected by the barrier yet,
whereas
electrons on the right ($x>0$) are. Note that in the case of perfect
reflection, $t=0$, we have $\tilde{J}_{3}=-J_{3}$ and the observable
densities (see below) form an Abelian $U(1)$ sub-algebra of $SU(2)$.
This is
the case of \textquotedblleft open boundary
bosonization\textquotedblright ,
which allows a complete and rather simple analysis. \cite{FabGo95} One
can
also show that the $J_{3}$ part of Eq.\ (\ref{3}) can be reduced by a
canonical transformation $H^{\prime }=U^{\dagger }HU$ to the Hamiltonian
with the interaction part
\begin{equation}
H_{1}^{\prime }=v_{F}\mathbf{B}.\mathbf{J}(x=0)-2g_{2}\int_{a}^{L}\upd %
x\,J_{3}(-x)J_{3}(x)\,,  \label{fictmagnfield}
\end{equation}%
here $U=\exp i\int_{0}^{\infty }\upd x\mathbf{B}.\mathbf{J}(x)$ and
$\mathbf{%
B}=2\theta (\cos \phi ,\sin \phi ,0)$. \cite{tobepub} The first term in
(\ref%
{fictmagnfield}) corresponds to the rotation of the incoming isospin
current
$\mathbf{J}$ by the ``magnetic field'',
$\mathbf{B}$, at the origin, Fig.\ \ref{fig:scattering}b. Eq.\
(\ref{fictmagnfield}) thus resembles the Hamiltonian for the Kondo
problem in the current algebra approach. \cite{AffLud91} The major
simplification in our
case is the classical nature of $\mathbf{B}$, as opposed to the quantum
Kondo spin $\mathbf{S}$, see \cite{AffLud91}.

\textit{Current and conductance}. The total electron density $\rho (x)$
is
given by
\begin{eqnarray}
\rho (x) &=&\left[ \rho _{iR}(x)+\rho _{oL}(x)\right]
 {\Theta} (-x)+
\left[%
\rho _{oR}(x)+\rho _{iL}(x)\right]
 {\Theta} (x),  \nonumber \\
&=&J_{0}(-x)+J_{0}(x)+\mbox{sgn}(x)\left[
-J_{3}(-|x|)+\tilde{J}_{3}(|x|)%
\right] \,,  \nonumber \\
&=&\rho _{c}(x)+\rho _{s}(x)\,,  \label{4}
\end{eqnarray}
where subscript $c(s)$ refers to isocharge (isospin) components. From
the
continuity equation $\partial _{t}\rho (x)=-\partial _{x}j(x)=-i[\rho
(x),H]$
one finds the current
\begin{eqnarray}
j(x) &=&v_{F}\left[
J_{0}(x)-J_{0}(-x)+J_{3}(-|x|)+\tilde{J}_{3}(|x|)\right]
\notag \\
&=&j_{c}(x)+j_{s}(x) \,.  \label{5}
\end{eqnarray}
We now consider the linear response to an applied voltage
$V(x,t)=\frac{1}{2}
V(t)\mbox{sgn}(x)$, which is seen to couple only to the isospin
components, $%
\rho _{s}$, of the density operator. The conductance is then given by
(in
units of $e^{2}/2\pi \hbar $)
\begin{equation}
G(x,t)=-2\pi i\Theta (t)\;\left\langle \left[ j_{s}(x,t)\;,\;
\int_{0}^{\infty }\upd y\;\rho _{s}(y,0)\right] \right\rangle \,,
\label{6}
\end{equation}
where we used the fact that correlation functions mixing the isocharge
and
isospin sectors vanish.

\bigskip

\textit{Perturbation theory in }$g_{2}$. The contributions to $G(\Omega
_{m}) $ in n-th order of $g_{2}$ may be calculated with the help of
Feynman
diagrams in the position-energy representation ($\Omega _{m}$ is the
external Matsubara frequency). We draw $n$ vertical wavy lines in
parallel,
the upper endpoint of the $i$-th line at $-x_{i}$ with isospin matrix $%
\frac12 \tau _{\alpha \beta }^{3}$ , the lower one at $x_{i}$ with
matrix $%
\frac12 R_{3\mu }{\tau }_{\alpha \beta }^{\mu }$ attached and carrying
the
factor $-2g_{2}$ ; $\alpha ,(\beta )$ are isospin indices of ingoing,
(outgoing) fermion lines. The external vertices are at $-x$ with matrix
$%
\frac12 \tau _{3}$ and at $y$ with matrix $\frac12 (R\vec{\tau})_{3}$.
The
vertex points are connected by Green's functions
\begin{equation}
\mathcal{G}(x\;;\;\omega _{n})=-\frac{i}{v_{F}} \mbox{sign}%
(\omega_{n})\Theta (\omega _{n}x) e^{-\omega _{n}x/v_{F}} \,,  \label{7}
\end{equation}
where the $\omega _{n}$ are Matsubara fequencies $\omega _{n}=(2n+1)\pi
T$.
All internal $x$ -variables are integrated on the positive semi axis.
The
trace over the product of all isospin matrices in each fermion loop is
taken
and a factor of $1/n!$ is applied to each n-th order diagram. The limit
$%
\Omega _{m}\rightarrow +0$ is taken at the end.

The incoming component of $\rho _{s}(y),J_{3}(-y)$, only contributes in
zeroth order: $G_{i}^{(0)}=\frac{1}{2}$. Adding the contribution from
the
outgoing component one finds
\begin{equation}
G^{(0)}=\frac{1}{2}(1+\cos 2\theta )=\cos ^{2}\theta =\left\vert
t\right\vert ^{2} \,.  \label{8}
\end{equation}

\begin{figure}[tb]
\includegraphics[width=8.5cm]{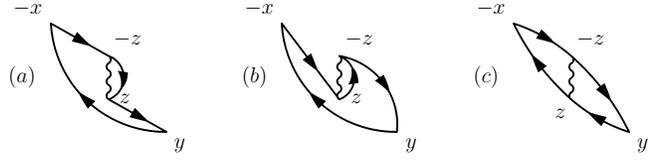}
\caption{ Three Feynman diagrams, depicting the first order in $g_{2}$
contribution to the conductance. The other three diagrams are obtained
from
these by reversing the direction of fermionic propagation. }
\label{fig:diag1order}
\end{figure}

The diagrams of first order in $g_{2}$ are shown in Fig.\ \ref%
{fig:diag1order}. Note that the \textquotedblleft vertex correction
type\textquotedblright\ diagram, Fig.\ \ref{fig:diag1order}c, is
$\propto
\Omega $ and will be dropped. One finds
\begin{equation}
G^{(1)}=-\frac{g_{2}}{4\pi }\sin ^{2}(2\theta )\;\ln \frac{T_{0}}{T}\,,
\label{9}
\end{equation}%
in agreement with \cite{Yue}. The ultraviolet cutoff, $T_{0}$, is
determined
by the width of the potential barrier $a$ as $T_{0}=v_{F}/(2\pi a)$, the
infrared cutoff arises at finite $T>v_{F}/L$ through the Green's
function at
time $t=0$: $\mathcal{G}(2a\;;\;t=0)\sim T/\sinh (2\pi Ta/v_{F})$. In
the
limit of zero temperature the finite-$T$ logarithm $\Lambda \equiv \ln
({%
T_{0}}/{T})$ is replaced by the zero temperature expression $\Lambda
_{0}\equiv \ln (L/a)$.

\begin{figure}[t]
\includegraphics[width=8.5cm]{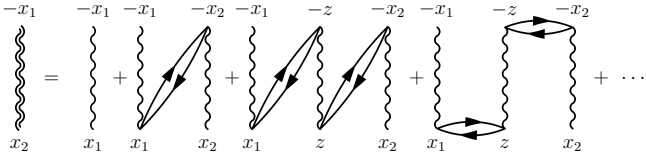}
\caption{Ladder series, $\bar{L}(x_{1},x_{2};\protect\omega _{n})$,
describing the combined effect of interaction and barrier, and leading
to a
linear-in-logarithm $\ln (T_{0}/T)$ contribution to conductance. }
\label{fig:ladder}
\end{figure}

In $n$-th order the diagrams with only one loop contribute the scale
dependent terms $[\ln T_{0}/T]^{n}$. Our prinicipal observation here is
that
the diagrams with the maximum number of loops ($n$ loops) contribute
linearly in logarithm $\propto \ln (T_{0}/T)$. They form a set of ladder
diagrams. The diagrams linear in $\ln (T_{0}/T)$ but not contained in
this
ladder series will be discussed \ below. The sum of all ladder diagrams
(see
Fig.\ \ref{fig:ladder}) may be calculated, and will be denoted by
$\bar{L}%
(x_{1},x_{2};\omega _{n})$. Later we will need the integrated quantity
$%
L(x_{1};\omega _{n})=\int_{0}^{\infty }\upd x_{2}e^{-\omega
_{n}x_{2}}\quad
\bar{L}(x_{1},x_{2};\omega _{n})$, which obeys the integral equation
\begin{eqnarray}
L(x;\omega ) &=&-ge^{-\omega x}\left[ 4\pi +\omega (Y+\frac{g}{2}%
)\int_{0}^{\infty }\upd ze^{-\omega z}L(z;\omega )\right]  \notag \\
&+&\frac{1}{2}g^{2}\omega \int_{0}^{\infty }\upd z\;\;e^{-\omega
|x-z|}L(z;\omega )\,.  \label{10}
\end{eqnarray}

\noindent Here we defined $Y=\cos 2\theta $, $g=g_{2}/2\pi $ and we use
units with $v_{F}=1$. The Eq.\ (\ref{10}) is of Wiener-Hopf type and its
solution is $L(x;\omega _{n})=-4\pi gbe^{-\omega _{n}px}$ where
$p=\sqrt{%
1-g^{2}}$ and $b=(1+p)/(1+p+gY)$. The conductance contribution $G^{(L)}$
in
linear order in $\ln \frac{T_{0}}{T}$, summed to all orders in $g_{2}$,
is
given by
\begin{eqnarray}
G^{(L)} &=&\frac{1-Y^{2}}{4}T^{2}\sum_{\epsilon ,\omega
}\int_{0}^{\infty }%
\upd x_{1}\upd x_{2}\upd y\bar{L}(x_{1},x_{2};\omega )\mathcal{G}%
(x_{1}-x;\epsilon )  \notag \\
&\times &\mathcal{G}(-x_{1}-x_{2};\epsilon -\omega )\mathcal{G}
(x_{2}-y;\epsilon )\mathcal{G}(y+x;\epsilon +\Omega ) \,.  \label{12}
\end{eqnarray}
Taking the limit $\Omega \rightarrow 0$ at $x\gg v_{F}/T$ one finds
\begin{equation}
G^{(L)}=\frac{-g(1-Y^{2})}{1+\sqrt{1-g^{2}}+gY}\;\ln \frac{T_{0}}{T}
\equiv
- \widetilde g(1-Y^{2})\ln \frac{T_{0}}{T} \,.  \label{13}
\end{equation}
Comparing eq.(\ref{13}) with eq.(\ref{9}) we see that the resummation
corresponds to dressing of the interaction, $g\to \widetilde g$.

\textit{Renormalization group approach}. In perturbation theory the
$n$-th
order contribution in $g_{2}$ is a polynomial in $\Lambda $ of degree
$n$.
If the theory is renormalizable, all terms with higher powers of
$\Lambda $
should be generated by a renormalization group equation for the scaled
conductance, $G(\Lambda )$, or equivalently for $Y(\Lambda )=2G(\Lambda
)-1$%
. Our approximation to the $\beta $-function of the RG equation is given
by
the prefactor of $\ln \frac{T_{0}}{T}$ in the perturbation theory result
(%
\ref{13}), with $Y$ replaced by $Y(\Lambda )$:
\begin{equation}
\frac{dY}{d\Lambda }=-\frac{2g(1-Y^{2})}{1+\sqrt{1-g^{2}}+gY}=\beta
_{L}(Y)\,.  \label{14}
\end{equation}%
Introducing the Luttinger parameter, $K=[(1-g)/(1+g)]^{1/2}$, we can
rewrite
(\ref{14}) as
\begin{equation}
\frac{d\Lambda
}{dY}=-\frac{1}{2(Y+1)(K^{-1}-1)}-\frac{1}{2(1-Y)(1-K)}\,,
\label{15}
\end{equation}%
which is easily integrated (see below). We note that (\ref{15}) has a
kind
of duality symmetry: it is invariant under $K\rightarrow K^{-1}$, $%
Y\rightarrow -Y$. For weak interaction, we expand $K^{\pm 1}\simeq 1\mp
g$
and recover the result by \cite{Yue}. In the limiting cases of nearly
transparent and nearly perfectly reflecting barrier, we recover the
results
by \cite{Kane,Furusaki}, with $G\sim \left( {T}/{T_{0}}\right)
^{2(\frac{1}{K%
}-1)}$ and $1-G\sim \left( {T_{0}}/{T}\right) ^{2(1-K)}$, respectively.

\begin{figure}[t]
\includegraphics[width=8.5cm]{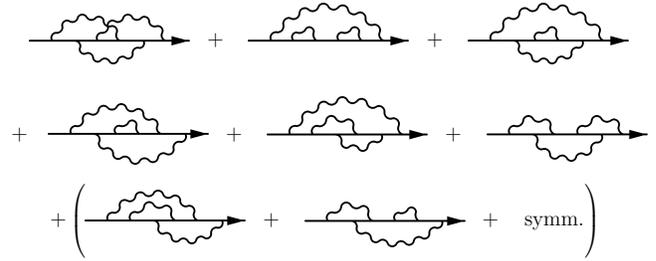}
\caption{Skeleton Feynman graphs, leading to lowest order logarithmic
contribution, $g^3 \Lambda$, beyond ladder series. }
\label{fig:3rd}
\end{figure}

A remaining question concerns the existence of additional terms in the
RG $%
\beta $-function, not contained in the ladder series (\ref{15}) (cf.\
\cite%
{LuWi03}). To answer it, we have calculated $G$ using computer algebra
up to
fourth order ($\sim 40,000$ diagrams). \cite{tobepub} The results are
summarized as follows. In higher orders we find both leading ($\sim
g^{n}\Lambda ^{n}$) and subleading ($\sim g^{n}\Lambda ^{k}$, $k<n$)
contributions. Most of these terms correspond to Eq.\ (\ref{14}).
However,
starting from third order of $g$ we find also subleading contributions
linear in $\Lambda $, which are explicitly different from the form
(\ref{15}%
) and arise from diagrams depicted in Fig.\ \ref{fig:3rd}. Adding these
terms to the $\beta $-function of (\ref{14}) we obtain
\begin{equation}
\frac{dY}{d\Lambda }=-\widetilde{g}%
(1-Y^{2})+g^{3}(1-Y^{2})^{2}(c_{3}-gYc_{4}+O(g^{2}))\,,  \label{corr}
\end{equation}%
with $\widetilde{g}$ the above result of the ladder resummation,
$c_{3}=\pi
^{2}/12$ and $c_{4}=0.238\ldots $. Notice the different power of
$(1-Y^{2})$
in front of the extra terms in (\ref{corr}), which renders these terms
irrelevant in the limits $G\rightarrow 0$, $G\rightarrow 1$.

Before proceeding further we should discuss the issue of universality of
the
logarithmic corrections, i.e.\ their dependence on the cutoff
regularization
scheme. Interpreting the ladder summation as a dressing of the
interaction,
Fig.\ \ref{fig:ladder}, $g\rightarrow \widetilde{g}$, the ladder result
(\ref%
{14}) thus corresponds to only one non-trivial Matsubara frequency
summation, which amounts to a "one-loop" RG correction in the usual
classification. In this case the scale invariant linear logarithmic
contribution is not sensitive to the cutoff regularization, i.e. when
$L$
exceeds the inverse temperature $\ln L/a\rightarrow \ln (v_{F}/2\pi Ta)$
and
the prefactor of the logarithm is preserved. In other words, these
contributions are universal in the RG sense. In the order $g^{2}$,
non-trivial "two-loop" corrections to (\ref{14}) are absent. In the
order $%
g^{3}$ the "three-loop" contributions are divided into two groups. The
first
group consists of the first diagram in Fig.\ \ref{fig:3rd} and its
symmetry-related partners: its contribution to $G$ is $\frac{\pi
^{2}}{24}%
\Lambda g^{3}(1-Y^{2})^{2}$ ; this linear-in-$\Lambda $ correction is
again
universal. The additional diagrams in Fig.\ \ref{fig:3rd} contain both
$%
\Lambda ^{3}$ and $\Lambda $ contributions. In this situation the
linear-in-$%
\Lambda $ terms contributing to the $\beta -$function are dependent on
the
cutoff scheme, i.e.\ are non-universal. The above cited value $c_{3}=\pi
^{2}/12$ corresponds to $T=0$ , i.e. the hard infrared cutoff $L$ in
real
space. If we calculate corrections at $T\gg v_{F}L^{-1}$, then we obtain
the
soft cutoff result $c_{3}=1/4$ instead, in agreement with \cite{LuWer07}
;
we use this latter value below.

Inverting (\ref{corr}) we get ${d\Lambda }/{dY}=(\beta
_{L}(Y))^{-1}-g\;h(g,Y)$ where $h(g,Y)=c_{3}-gYO(1)+\ldots $ for small
$g$
has been defined. Notice that in the non-ladder corrections $g\;h(g,Y)$
we
may substitute $g$ by its renormalized value $g\rightarrow
\widetilde{g}$.
Truncating at lowest level beyond the ladder series, we get
\begin{equation}
{d\Lambda }/{dY}\simeq
\left[ {\beta _{L}(Y)}\right]^{-1}-c_{3}\widetilde{g}
\;\,.  \label{corr2}
\end{equation}
Integrating the latter equation we have
\begin{eqnarray}
\left( {T}/{T_{0}}\right) ^{2(1-K)}
&=& {\Phi(G)}/{\Phi (G_{0})} \,,\label{bestapprox}
\\
\Phi(G)&=&\frac{G^{K}}{1-G}
\left( K+G(1-K)\right) ^{4c_{3}(1-K)}\,,
\nonumber
\end{eqnarray}
with $c_{3}=1/4$ and the assumed initial condition $G=G_{0}$ at
$T=T_{0}$.

The equation (\ref{bestapprox}) is the central result of this paper. Let
us
discuss it in more detail. The above ladder approximation(\ref{14})
would
correspond to setting $c_{3}=0$ in (\ref{bestapprox}). It is seen that
the
above cited scaling-law dependences of $G$ on $T$ remain asymptotically
exact at $G\rightarrow 0$ and $G\rightarrow 1$. The existence of terms
in
the $\beta $-function beyond the ladder series modifies the behavior of
the
conductance at intermediate values $G\sim 1/2$, and the role of
$h(g,Y)\sim
1/4$ can hence be largely viewed as a \emph{redefinition of the cutoff
energy%
} $\ln T_{0}\rightarrow \ln T_{0}+O(gG_0)$, when going from higher to
lower $%
T$. The duality symmetry, Eq.\ (\ref{15}), between the scaling exponents
is
preserved, $T^{2(K-1)} \to T^{2(K^{-1}-1)}$, in contrast to the recent
claim
in \cite{Enss05} ; the breaking of duality reported in \cite{Enss05}
might
be connected to the approximate character of the solution for the set of
flow equations there.

Let us also compare our findings to exact expressions available from the
thermodynamic Bethe Ansatz method. In the particular case of $K={1}/{2}$
($g=3/5$) the conductance $G$ is obtained as a closed analytic function
of $T$. \cite{Kane,Weiss} Our Eq.\ (\ref{bestapprox}) reduces to a
quadratic equation, with relevant solution
\begin{equation}
G=\frac{4\tau ^{2}}{1+4\tau ^{2}+\sqrt{1+16\tau ^{2}}}\,,\quad \tau
=\frac{T%
}{T_{0}}\frac{\left\vert t\right\vert \sqrt{1-\frac{1}{2}|r|^{2}}} {%
\left\vert r\right\vert ^{2}}\,,  \label{19}
\end{equation}%
where $\left\vert t\right\vert $, $\left\vert r\right\vert $ are the
bare
transmission and reflection amplitudes.

\begin{figure}[t]
\includegraphics[width=\columnwidth]{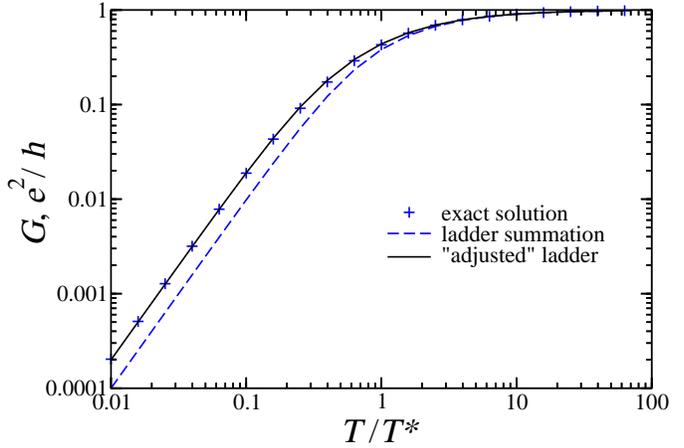}
\caption{Conductance as a function of temperature for $K=1/2$. The exact
result available from TBA (multiplied by factor 2) is shown by pluses.
\protect\cite{Kane,Weiss} The results of ladder summation,
(\protect\ref%
{bestapprox}) with $c_3=0$, are shown by dashed line ; the result of
Eq.\ (%
\protect\ref{bestapprox}) with the adjusting term $c_3=1/4$ is shown by
solid line. }
\label{fig:conduc}
\end{figure}

At high temperatures the result reported in \cite{Kane,Weiss} tends to
the
clean limit value for a four-point measurement, $G=\frac{1}{2}$; here we
assume that the two-point conductance is obtained by multiplying the
latter
result by a factor of $1/K$ equal to $2$ in order to make contact with
our
theory. Fixing the overall scale $T^{\ast}$ in both solutions
($\tau \equiv {T}/{T^{\ast}}$) by their high-temperature behavior,
$G\simeq 1-\tau ^{-1}$,
we show the overall picture in Fig.\ \ref{fig:conduc}. It is seen that
the
result of the ladder summation overestimates the renormalization of the
barrier, predicting smaller conductance at low $T$. At the same time the
adjustment of the ladder summation by $c_{3}=1/4$, Eqs.(\ref{corr2}),
(\ref{bestapprox}), (\ref{19}), gives excellent agreement with the exact
solution
\cite{Kane,Weiss}, with a relative deviation not exceeding 4 \% in the
whole
temperature range.

 {In case $K={1}/{3}$ ($g=4/5$), relevant for the description of
point conductance between the quantum Hall edge states, \cite{Fendley95}
we do not have a closed analytic expression for $G(T)$. It is however
possible to make a comparison to our theory. We fix the overall
temperature scale by adopting $G \simeq 1-\tau ^{-4/3}$ at high
temperatures, then we have  $G \simeq 9 \tau ^{4}$ from our Eq.\
(\ref{bestapprox}).
Multiplying again
the result reported in \cite{Fendley95} by a factor of $1/K = 3$,
we have  $G \simeq 10.0638 \tau ^{4}$ at low $T$,
whose prefactor is within 12 \% from our value. }
 We may thus conclude that our result
(\ref{bestapprox})
provides a very good approximation even in the strongly interacting case
$g\sim 1$, i.e.\ it is sufficient for all practical purposes.

\textit{Conclusion.} In this letter we presented a theory of transport
of
interacting electrons through a potential barrier, in the linear
response
regime and at all temperatures, for any short range barrier and for any
forward scattering interaction. We employed a representation in terms of
chiral fermions, which greatly simplifies the perturbation theory in the
interaction parameter $g_{2}$. In this way the scale dependent
contributions
to the conductance may be studied systematically. At low energies long
range
correlations lead to logarithmically diverging terms in the form of
powers of $\Lambda =\ln \frac{T_{0}}{T}$.

In particular, the terms linear in $\Lambda $ may be summed to all
orders in
$g_{2}$, and the prefactor may be identified with the $\beta$-function
of the renormalization group equation for the conductance as a function
of the scaling variable $\Lambda$. At intermediate temperatures
additional small corrections to the $\beta$-function were found in third
and higher orders of perturbation theory. Approximating these additional
terms by the lowest order (in $g_{2}$ ) gave excellent agreement with
the known exact result at $K=\frac{1}{2}$. This appears to be one of the
few cases where the $\beta$-function can be determined beyond
perturbation theory. Our results are in
agreement with all known results, where applicable, but go far beyond.
The
RG-equation may be integrated analytically to give the conductance as an
implicit function of the temperature. The method we describe here is
quite
general and may be of value for calculating transport properties of the
model out of equilibrium or of other models in which the
$\beta$-function
may be obtained by summing a ladder series.

\acknowledgments

We are grateful to I.V.\ Gornyi, D.G.\ Polyakov, K.A.\ Matveev, D.A.\
Bagrets, O.M.\ Yevtushenko, M.N.\ Kiselev, A.M.\ Finkel'stein, A.A.\
Nersesyan, L.I.\ Glazman, S. Lukyanov for various useful discussions. PW
acknowledges the Aspen Center for Physics, where part of this work has
been
performed.

%\begin{figure} \includegraphics{epl-template.eps}
%\caption{Figure caption.}  \label{fig.1} \end{figure}


\begin{thebibliography}{99}
\bibitem{Apel} \Name{Apel W. \and  Rice T. M.}
\REVIEW {Phys. Rev. B}{26} {1982} {7063}.

\bibitem{Kane} \Name {Kane C. L. \and Fisher M. P. A.}
\REVIEW {Phys.
Rev. Lett.}{68} {1992} {1220} ; \REVIEW {Phys. Rev. B}{46} {1992}
{15233}.

\bibitem{Maslov} \Name {Maslov D. L. \and  Stone M.}
\REVIEW {Phys. Rev. B}{52} {1995} {R5539}.

\bibitem{Safi} \Name {Safi I. \and Schulz H. J.}
\REVIEW {Phys. Rev. B}{52} {1995} {R17040}.

\bibitem{Oreg} \Name {Oreg Y. \and Finkel'stein A. M.}
\REVIEW {Phys. Rev. B}{54} {1996} {14265}

\bibitem{Furusaki} \Name {Furusaki A. \and  Nagaosa N.}
\REVIEW {Phys. Rev. B}{47} {1993} {4631}.

\bibitem{Weiss} \Name {Weiss U., Egger R. \and Sassetti M.} \REVIEW
{Phys. Rev. B}{52} {1995} {16707}.

 {
\bibitem{Fendley95}
\Name{ Fendley P., Ludwig A. W. W.  \and  Saleur H.}
\REVIEW{Phys. Rev. Lett.}{74}{1995}{3005}; \REVIEW{Phys.
Rev. B}{52}{1995} {8934}. }

\bibitem{Yue} \Name {Yue D., Glazman L. I. \and  Matveev K. A.} \REVIEW
{Phys. Rev. B}{49} {1994} {1966}.

\bibitem{Gornyi} \Name {Polyakov D. G. \and  Gornyi I. V.} \REVIEW
{Phys. Rev. B} {68} {2003} {035421}.

\bibitem{GoNeTs} \Name {Gogolin A. O., Nersesyan A. A. \and Tsvelik A.
M.} \Book {Bosonization and Strongly Correlated Systems}
\Publ {Cambridge
University Press, Cambridge} \Year {1998}.

\bibitem{AffLud91} \Name { Affleck I.} \REVIEW { Nucl. Phys. B}{336}
{1990} {%
517} ; \Name { Affleck I. \and  Ludwig A. W. W.}
\REVIEW {
Nucl.
Phys. B}{360} {1991} {641}.

\bibitem{FabGo95} \Name { Fabrizio M. \and  Gogolin A.O.} \REVIEW
{Phys. Rev. B}{51} {1995} {17827}.

\bibitem{tobepub} \Name { Aristov D. N. \and  W\"olfle P.} to be
published.

\bibitem{LuWi03} \Name { Ludwig A. W. W. \and  Wiese K. J.}
\REVIEW {
Nucl. Phys. B} {661} {2003} {577}.

\bibitem{LuWer07} \Name { Lukyanov S. L. \and  Werner Ph.}
\REVIEW {J.
Stat. Mech.}{ } {2007} {P06002}; \Name {Lukyanov S. L.} private
communication.

\bibitem{Enss05}
\Name{Enss T., Meden V., Andergassen S.,
Barnab\'e-Th\'eriault X., Metzner W., \and Sch\"onhammer K.} \REVIEW {Phys.
Rev. B}{71} {2005} {155401}.


\end{thebibliography}
\end{document}